\documentclass{raa}
\usepackage{graphicx,times}

\begin{document}

   \title{SEARCH FOR POSSIBLE EXOMOONS WITH FAST TELESCOPE}

 \volnopage{ {\bf 2017} Vol.\ {\bf X} No. {\bf XX}, 000--000}
   \setcounter{page}{1}

   \author{Dragan V. Luki{\' c}}

   \institute{Institute of Physics, University of Belgrade,
	P.O.Box 57, 11001 Belgrade, Serbia; {\it lukic@ipb.ac.rs}}


   \date{\small Received 2012 June 12; accepted 2012 July 27}

\abstract{Our knowledge of the Solar System, encourages us to beleive that we might expect exomoons to be present around some of the known exoplanets. With present hardware and 
existing optical astronomy methods we shall not be able to find exomoons at least 10 years from now and even then, it will be a hard task to detect them. Using data from the Exoplanet Orbit 
Database (EOD) we find stars with Jovian exoplanets within 50 light years. Most of them will be fully accessible by the new radio telescope, The Five-hundred-meter Aperture Spherical 
radio Telescope (FAST) under construction, now in the test phase. We suggest radio astronomy based methods to search for possible exomoons around two exoplanets.
 }

   \authorrunning{Dragan V. Luki{\' c}}            
   \titlerunning{SEARCH FOR POSSIBLE EXOMOONS WITH FAST TELESCOPE}  
   \maketitle



%
\section{Introduction}           
\label{sect:intro}
Discovery of the 51 Pegasi b (Mayor \& Queloz \cite{Mayor95}), exoplanet orbiting the Sun-like main sequence star, was only the first in the series of many exoplanets discovered thenceforth. The 
progress is made thanks to advanced detection techniques and instrumentation. Nowadays, the result is thousands of confirmed and potential exoplanets, the most of them 
identified by the NASA's Kepler space telescope. Every prudent connoisseur of the Solar System would expect the presence of the exomoons close to the known exoplanets. The Solar 
System's planets and dwarf planets are known to be orbited by 182 natural satellites. In our solar system, Jovian planets have the biggest collections of moons and we expect 
similar position for gas giant planets in extrasolar systems. Nonetheless, contemporary techniques for observation haven't made a single detection of any exomoon so far.

One of the leading models describing planetary satellites formation is the actively supplied gaseous accretion disk model (Canup \& Ward  \cite{Canup06}). In this model, the final 
total mass of satellite system, approximately $10^{-4}$ $M_P$ ($M_P$ mass of planet) is given by a balance of the supply of material to the satellites, and satellite 
loss through orbital decay driven by the gas. Dust grains stick and grow to form satellitesimals within a circumplanetary disk. An alternative model is the 
solids enhanced minimum mass model (see e.g. Masqueira \& Estrada \cite{Masqueira03}). The model only qualitatively describes the expected mass ratios. Since this model does not give 
us the masses of satellites we will use only the first model for our analysis of possible exomoons.

\section{SELECTION OF DATA AND METHOD OF ANALYSIS}
\label{sect:Sel}

Many of the detected exoplanets are the gas giants located in the habitable zone of their stars. These big planets cannot support life, but it is believed that some of their exomoons 
could be habitable. In our analysis, assuming that scaling law (Canup \& Ward \cite{Canup06}) observed in the solar system also applies for extrasolar super-Jupiters 
(Heller \& Pudritz \cite{Heller14}), we used  planet's data from the Exoplanet Orbit Database catalog (Wright et al. \cite{Wright11}, and Han et al. \cite{Han14}). We selected only exoplanets closer 
than 50 light years which have comparable mass or are more massive than Jupiter within declination limits of full sensitivity of the new radio telescope in China. Approximately a half of 
them are hot or warm Jupiters. According to (Heller \& Pudritz \cite{Heller15}) if these planets migrated into the stellar habitable zones from beyond a few AU, they could be orbited 
by large, water rich satellites. The liquid water on the surface is possible on sufficiently massive satellites.

Besides telescopes explained in (Griessmeier et al. \cite{Griessmeier11}) and (Noyola \cite{Noyola15}) we have additional radio telescope in the final phase of construction, The Five-hundred-meter 
Aperture Spherical radio Telescope (FAST) (Nan et al. \cite{Nan11}) and latter SKA telescope. FAST is located at a great depression with a diameter of about 800 m at 25$^{o}$.647N and 106$^{o}$.856E, near 
the village of Dawodang, in Guizhou Province, China. FAST will be capable of covering the sky within 40$^{o}$ from the zenith with full sensitivity. Set of nine receivers covers a frequency range from 70 MHz to 3 GHz. 
It has an illuminated aperture of 300 m diameter within 500 m diameter of reflector. FAST is an order of magnitude more sensitive than 100-m telescopes at Green Bank, USA, and Effelsberg, Germany and about two times more than 
Giant Meterwave Radio Telescope (GMRT), India. 

They will try the first direct detection of exoplanets in meter wave band (Li et al. \cite{Li12}). The quasi-periodicity of planetary radio burst is tied to the spin 
of the planet, which is in the order of days. This time-modulation of radio signals augument the detectability of exoplanets by FAST. The idea of direct radio detection of exoplanets was first suggested by 
(Lecacheux \cite{Lecacheux90}). For the search for exomoons we count on interaction of magnetic field of extrasolar planets with plasmas from exomoons (Zarka \cite{Zarka07}). In Solar system, Jupiter's magnetosphere radiate intense decameter 
radio waves. From the Earth, these radio waves are detectable in the frequency range from 10 to 40 MHz (Zarka \cite{Zarka98}). The generation mechanism is the cyclotron-maser instability (Wu \&Li \cite{Wu79}). Selected 
exoplanets are presented in Table 1. Listed frequencies in Table 1 are the maximum values reported in (Griessmeier et al. \cite{Griessmeier07}). In Table 2 we compare frequencies $f_{mG}$ calculated by 
(Griessmeier et al. \cite{Griessmeier07}), with other models of extrasolar planets radio emission $f_{mL}$ (Lazio et al.\cite{Lazio04}) and $f_{mR}$ (Reiners \& Christiansen \cite{Reiners10}) and expected radio fluxes. Other 
radio telescopes with the most suitable frequency range are: super LOFAR extension (NenuFAR, 10-80 MHz) in France, 1-2 mJy at 4MHz bandwidth, Giant Meterwave Radio Telescope (GMRT, 153 MHz) in India, 0.2 mJy/sqrt(t/15 minutes) 
in a 4 MHz bandwith, Ukrainian T-shaped Radio telescope (UTR-2, 8-40 MHz), and Arecibo (47 Mhz).

We can see that the selected planets in Table 1 orbit stars from M to F star type. The closest planets are eps Eridani b at 3.22 pc and Gliese 876 b and c at 4.69 pc distance. Since these planets most likely 
were not formed at those distances, but have migrated from larger ones, possible exomoons could also be the captured rocky planets. Now if the possible exomoons are captured they can survive enough time for 
all stars presented in Table 1 (Barnes \& Brien \cite{Barnes02}). These captured satellites can be more massive than the formed ones (Porter \& Grundy \cite{Porter11}, and Teachey et al. \cite{Teachey17}). Since we are not 
aware of theory which can predict such an event we do not consider them. Even if 50 percent planets are falsely detected (Santerne et al. \cite{Santerne15}) we still have enough candidates. We will have most likely high mean 
plasma density between $\rho$S$\sim$ 10$^6 $amu cm$^{-3}$ and $\rho$S$\sim$ 10$^7 $amu cm$^{-3}$ due to the presence of some exomoons in star's habitable zone and closer to stars (Schunk \& Nagy \cite{Schunk09}).

\section{SOME OF THE CLOSEST EXOPLANETS WITH POSSIBLE EXOMOONS}

All stars fully accessible by the FAST are: eps Eridani, Gliese 876, Gliese 849, HD 62509, 55 Cancri, HD 147513, Upsilon And A, 47 UMa b, HIP 79431 and HD 176051, and have enough lifetime to be listed in HabCat (Turnbull \& Tarter \cite{Turnbull03}).
If, as we can see in Table 1, these stars do not have exomoon emitters with frequency above 70 MHz our next chance is the low-end Low-Frequency Array (LOFAR). Present LOFAR has frequency range 10$-$240 MHz which is the best for 
exomoons and exoplanets detection. The super LOFAR extension (NenuFAR, 10-80 MHz) has frequency range of our interest. Sensitivity in this range is a few mJy. Let us check some of the closest planets from our list around stars 
eps Eridani, Gliese 876, Gliese 849, 55 Cancri, and Upsilon And A.

Eps Eridani is an orange dwarf star located 3.22 pc away. Exoplanet Eps Eridani b is a prime target for future extrasolar planet direct-imaging attempt due to its proximity. Its mass 
was calculated to be 0.83 $M_{\odot}$. It was target in the previous measurements (George \& Stevens \cite{George08}, and Noyola \cite{Noyola15}) with GMRT. They did not find signs of 
exomoon radio activity. It is fully accessible with FAST. NenuFAR would be good for search for radio emission of possible exomoons around this star.

Gliese 876 is a red dwarf star located 4.69 pc away. It has two larger exoplanets Gliese 876 b (2.27$M_{\odot}$) and Gliese 876 c (0.7$M_{\odot}$). Both planets lie in the habitability 
zone around the star. It is fully accessible with FAST and the GMRT but planets' maximum frequencies are below frequencies of these telescopes which leaves only NenuFAR telescope. 

Gliese 849 is a red dwarf at 9.1 pc away from the Sun. It has big separation of planets from star. The orbital separation of Gliese 849 b (2.39 AU) amounts to the angular
separation of 0.25$''$. Due to the proximity, it provides a good chance for high-resolution imaging using adaptive optics. It is fully accessible with FAST and the GMRT. Maximum 
frequency of Gliese 849 b is 21.8 MHz much below frequencies of these telescopes beside NenuFAR.

55 Cancri is a binary star 12.3 pc away from the Sun. The system consists of a yellow dwarf star 55 Cancri A, and a smaller red dwarf 55 Cancri B. The primary star, 55 Cancri A  is more 
enriched than the Sun in elements heavier than helium, with 186$\%$ the solar abundance of iron. It is classified as a rare "super metal-rich" (SMR) (Marcy et al. \cite{Marcy02}). It has only a low 
emission from its chromosphere. The 55 Cancri A system has at least five planets. Exoplanet 55 Cancri b is a hot Jupiter and it has 0.95 $M_{J}$ mass.  Due to the vicinity of the star, tidal 
forces would either eject exomoons from orbit or destroy them, so it is not expected to have large ones (Barnes \& Brien \cite{Barnes02}). 55 Cancri d is orbiting on distance 5.74 AU and it has 3.878
$M_{J}$ mass.  Maximum calculated frequency of 55 Cancri d is close to 70 MHz (Griessmeier et al. \cite{Griessmeier07}). It is fully accessible with FAST and the GMRT.

\noindent
\parbox{\textwidth}{
\scriptsize
{\bf Table 1.} Possible exomoons
\vskip.25cm \centerline{
		\rm\begin{tabular}{|l|l|l|l|l|l|l|l|}
			\hline
			$Planet\, Name$ & $Mass$   &$Star\, type$ &$Semimajor$ & $Distance$ & $Satellite$ & $Declination$ & Frequency\\
			& & &$Axis$ &  [pc] & $mass$ &  & \\
          		\hline
			eps Eridani b&  1.55 M$_{J}$ & K2V  & 3.4 AU & 3.22 & 0.049 M$_{\oplus}$ &-09$^{o}$ 27$'$ 29.7312$''$ & 33.2 MHz\\
			Gliese 876 b &  2.27 M$_{J}$ & M4V & 0.2 AU & 4.69 & 0.072 M$_{\oplus}$ &-14$^{o}$ 15$'$ 49.32$''$ & 38.2 MHz\\	
			Gliese 876 c &  0.7 M$_{J}$ & M4V & 0.13 AU & 4.69 & 0.022 M$_{\oplus}$ &-14$^{o}$ 15$'$ 49.32$''$ & 16.4 MHz\\
			Gliese 849 b &  0.91 M$_{J}$ & M3.5V & 2.39 AU & 9.1 & 0.0289 M$_{\oplus}$ &-4$^{o}$ 38$'$ 26.62$''$& 21.8  MHz\\
			Gliese 849 c &  0.94 M$_{J}$ & M3.5V & 4.82 AU & 9.1 & 0.0298 M$_{\oplus}$ &-4$^{o}$ 38$'$ 26.62$''$&  \\
			HD 62509 b &  2.9 M$_{J}$ & K0III & 1.64 AU & 10.3 & 0.092 M$_{\oplus}$ &+28$^{o}$ 01$'$ 35$''$ & 49.5 MHz\\
			55 Cnc b &  0.8 M$_{J}$ & G8V & 0.11 AU& 12.3 & 0.025 M$_{\oplus}$ &+28$^{o}$ 19$'$ 51$''$ & 18.9 MHz\\
			55 Cnc d &  3.878 M$_{J}$ & G8V  & 5.74 AU & 12.3 & 0.123 M$_{\oplus}$ &+28$^{o}$ 19$'$ 51$''$ & 61.4 MHz\\
			HD 147513 b &  1.21 M$_{J}$ & G1VH-04 & 1.32 AU & 12.9 & 0.038 M$_{\oplus}$&+39$^{o}$ 11$'$ 34.7121$''$ & 24.5 MHz\\
			ups And A b   & 0.62 M$_{J}$ & F8V & 0.059 AU& 13.47 &  0.019 M$_{\oplus}$&+41$^{o}$ 24$'$ 19.6443$''$& 2.4 	MHz\\
			ups And A c   & 13.98 M$_{J}$ & F8V & 0.832 AU& 13.47 &  0.44 M$_{\oplus}$&+41$^{o}$ 24$'$ 19.6443$''$ & 38.4	MHz\\
			ups And A d   & 10.25 M$_{J}$ & F8V & 2.53 AU& 13.47 &  0.33 M$_{\oplus}$&+41$^{o}$ 24$'$ 19.6443$''$ & 61.4	MHz\\
			ups And A e   & 0.96 M$_{J}$ & F8V & 5.25 AU& 13.470 &  0.031 M$_{\oplus}$&+41$^{o}$ 24$'$ 19.6443$''$ &	\\
			47 UMa b &  2.5 M$_{J}$ & G1V & 2.1 AU  & 14.06 & 0.079 M$_{\oplus}$&+40$^{o}$ 25$'$ 27.97$''$ & 46.6 MHz\\
			HIP 79431 b &  2.00  M$_{J}$ & M3V & 0.36 AU & 14.4  & 0.064 M$_{\oplus}$&-18$^{o}$ 52$'$ 31.8$''$&  40 MHz \\
			HD 176051 b & 1.5 M$_{J}$ & F9V & 1.76 AU & 15 & 0.047 M$_{\oplus}$&+32$^{o}$ 54$'$ 5$''$ & \\
			\hline
		\end{tabular}
} } \vskip .5cm

Upsilon Andromedae is a binary system. The system consists of a yellow-white dwarf star $\upsilon$ And A, and of a red dwarf star $\upsilon$ And B, (Lowrance \& et al.\cite{Lowrance02}, and Santos et al. \cite{Santos04}). 
Declination of system is +41$^{o}$ 24$'$ 19.6443$''$ (van Leeuwen \cite{van Leeuwen07}). It is fully accessible with FAST and the GMRT. The separation between the stars is 750 AU (Lowrance et al. \cite{Lowrance02}). 
Radial-velocity measurements led to the detection of four planets around $\upsilon$ And A, and one of the planets, i.e., $\upsilon$ And A d, is found to be located within $\upsilon$ And A's 
habitable zone. The planet stays inside the extended zones of habitability at its apoapsis at 3.26 AU. Its periapsis is given as 1.76 AU, a distance close to the inner limit of the general 
habitability zone. General and extended zones of habitability, are in use as in (Kasting et al. \cite{Kasting93}) and subsequent works.  The planets $\upsilon$ And A b, c, d and e have the semi-major 
axes as 0.0592, 0.828, 2.51, and 5.25 AU. Measured mass of planet $\upsilon$ And A d as 3.75 $M_{J}$(Ligi et al\cite{Ligi12}). The eccentricity of $\upsilon$ And A d is identified as 
0.299 (Curiel et al. \cite{Curiel11}), and its true mass has been estimated as 10.19 $M_{J}$ (Barnes et al. \cite{Barnes11}) and 10.25 $M_{J}$ (McArthur et al. \cite{McArthur10}). The mutual inclination between the planets 
c and d, is large as 30$^{o}$ (Barnes et al.\cite{Barnes11}). $\upsilon$ And A is the only multiplanetary system with astrometry measurements (Deitrick et al. \cite{Deitrick15}). The age of the star is about 
3 Gyr (McArthur et al. \cite{McArthur10}, and Takeda et al. \cite{Takeda07}).  Maximum calculated radio frequency of $\upsilon$ And A d is close to 70 MHz 
(Griessmeier et al. \cite{Griessmeier07}), the lowest frequency range of receiver at FAST similar to the 55 Cancri d.

Gliese 86 A (13 G. Eridani) is an orange dwarf main-sequence star with -50$^{o}$ 49$'$ 25.4179$''$ declination of system (van Leeuwen \cite{van Leeuwen07}) approximately 
10.8 pc away in the southern constellation Eridanus. Its binary companion is Gliese 86 B, a white dwarf star. Exoplanet Gliese 86 Ab (Table 2) is the most promising candidate 
but it lies out of the field of view of full sensitivity of all radiotelescopes mentioned in this paper and it needs to wait for SKA telescope. Gliese 86 A has lower metallicity 
than our Sun unlike the most stars with exoplanet. The orbit of the planet is almost circular, it has an eccentricity of 0.05, and a period of 15.83 days. These planet's 
observed characteristics, combined with double-star nature, suggest that planetary systems maybe was not formed in the standard agglomeration scheme.

We can see that the most suitable radio telescope for search for the closest possible exomoons is NenuFAR, the super LOFAR extension and FAST telescope especially for two 
extrasolar planets, 55 Cancri d and $\upsilon$ And A d, where it can be very useful. The both possible exomoons fulfill requirement to be more massive than Mars (Heller \& Pudritz  \cite{Heller15}). 
As we can see in Table 2. for other models of extrasolar planets radio emission (Lazio et al. \cite{Lazio04}, and Reiners \& Christiansen \cite{Reiners10}) second set of FAST receivers (Nan et al. \cite{Nan11}) 
is also suitable for these two extrasolar planets and plans for radioastronomy methods search for exoplanet (Li et al. \cite{Li12}). To our best knowledge we do not know for 
the exoplanet detection by radioastronomy methods. 

\vskip.5cm{}

\noindent
\parbox{\textwidth}{
	{\bf Table 2.} Possible maximum frequencies for exomoons and exoplanets 
	\vskip.25cm \centerline{
		\rm\begin{tabular}{|l|l|l|l|l|l|l|l|l|}
			\hline
			$Planet\, Name$  &$Star\, type$ &$SemiMajor$ & $f_{mG}$ &R. flux&$f_{mL}$ &R. flux& $f_{mR}$& R. flux\\
			&[M$_{J}$] &$Axis$   & [MHz]&[mJy]& [MHz]&[mJy]&[MHz]&[mJy]\\
			\hline
			eps Eridani b& K2V  & 3.4 AU &  33.2 &0& 53& 6.3&18.3&6\\
			Gliese 876 b & M4V & 0.2 AU & 38.2& 6.3& 66&3.1&68&160\\	
			Gliese 876 c & M4V & 0.13 AU &  16.4 &61.7& 16& 2.1&8.9&630\\
			Gliese 849 b & M3.5V & 2.39 AU & 21.8&0&&&&\\
			Gliese 849 c & M3.5V & 4.82 AU  & & &&&&\\
			HD 62509 b   & K0III & 1.64 AU  & 49.5 &0& 68&0.1&&\\
			Gliese 86 Ab  & K1V & 0.11 AU & 61 &3.8 & 113& 43.8&237&63\\
			55 Cnc b     & G8V & 0.11 AU&  18.9&3& &&17.6&80\\
			55 Cnc d     & G8V  & 5.74 AU &61.4 &0 && &242&0\\
			HD 147513 b  & G1VH-04 & 1.32 AU &24.5&2& 43&4.1&23.5&0.2\\
			ups And A b  & F8V & 0.059 AU& 2.4&178.5  & 27 &41.8&2.2&200\\
			ups And A c  & F8V & 0.832 AU& 38.4&0& 84&2.8&68&2.5\\
			ups And A d  & F8V & 2.53 AU&  61.4&0	& 163&0.1&213&0.3\\
			ups And A e  & F8V & 5.25 AU&   &&	&&&\\
			47 UMa b &  G1V & 2.1 AU   & 46.6& 0&&&111&0.5\\
			HIP 79431 b & M3V & 0.36 AU   & 40&0& &&&\\
			HD 176051 b &  F9V & 1.76 AU & &  & &&&\\
			\hline
		\end{tabular}
	} }

\section{Conclusion}
\label{sect:conclusion}
Since we shall not be able to find exomoons with existing optical astronomy methods at least 10 years from now (Kipping  \cite{Kipping14}) and even then it will be hard task to detect them
(Hippke \& Angerhausen \cite{Hippke15}, and Heller et al. \cite{Heller16}) we suggest to search for exomoons around these planets with radio astronomy based methods 
(see Noyola et al. \cite{Noyola14}, and Noyola \cite{Noyola15}). The main problem could be that the distances of the exoplanets we are suggesting for investigation with the FAST 
telescope are greater than the ones selected in the first searches (George \& Stevens \cite{George08}, and Noyola \cite{Noyola15}). The closest planets are 55 Cancri d at 12.3 pc 
and $\upsilon$ And A d at a distance of 13.47 pc. At present such sensitivity can be expected only from the radio telescope the super LOFAR extension (NenuFAR, 10-80 MHz) and the just finished radio telescope FAST.

\normalem
\begin{acknowledgements}
This research has been supported by the Ministry of Education and Science of the Republic of Serbia through the project: 176021 'Visible and Invisible Matter in Nearby Galaxies: Theory and observations'. 
This research has made use of the Exoplanet Orbit Database and the Exoplanet Data Explorer at exoplanets.org.
\end{acknowledgements}


\label{lastpage}

\end{document}